\documentclass[useAMS,usenatbib]{mn2e}
\usepackage{graphicx}
\title[Merger rates from merger trees in the extended Press-Schechter theory]
{Merger rates of dark matter haloes from merger trees in the
extended Press-Schechter theory}
\author[Nicos  Hiotelis]{Nicos ~Hiotelis\thanks{E-mail:
hiotelis@ipta.demokritos.gr} \thanks{Present address: Lysimahias
66, Neos
Kosmos, Athens, 11744 Greece}\\
1st Experimental Lyceum of Athens, Ipitou 15, Plaka, 10557,
Athens, Greece, E-mail: hiotelis@ipta.demokritos.gr}
\begin{document}

\date{Accepted ............... Received ................; in original form ...........}

\pagerange{\pageref{firstpage}--\pageref{lastpage}} \pubyear{2008}

\maketitle

\label{firstpage}

\begin{abstract}
We construct merger trees based on the extended Press-Schechter
theory (EPS) in order to study the merger rates of dark matter
haloes over a range of present day mass ($10^{10}M_{\sun}\leq M_0
\leq10^{15}M_{\sun}$), progenitor mass $(5\times10^{-3}\leq \xi
\leq1$) and redshift ($0\leq z\leq 3$). We used the first crossing
distribution of a moving barrier of the form
$B(S,z)=p(z)+q(z)S^{\gamma}$, proposed by Sheth \& Tormen,  to
take into account the ellipsoidal nature of collapse. We find that
the mean merger rate per halo $B_m/n$ depends on the halo mass $M$
as $M^{0.2}$ and on the redshift as
$(\mathrm{d}\delta_c(z)/\mathrm{d}z)^{1.1}$. Our results are in
agreement with the predictions of N-body simulations and this
shows the ability of merger-trees based on EPS theory to follow
with a satisfactory agreement the results of N-body simulations and
the evolution of structures in a hierarchical Universe.
\end{abstract}

\begin{keywords}
   cosmology: theory -- dark matter, galaxies: haloes -- structure -- formation, methods: analytical
\end{keywords}
%
%________________________________________________________________

\section{Introduction}

Although numerical experiments are the most powerful methods to
study the formation of structures, the development of analytical
or semi-numerical methods is very important, since they help to
improve our understanding of the physical processes during the
formation. \\
A class of analytical methods is that based on the ideas of Press
\& Schechter (1974) and on their extensions (Bond et al.
1991, Lacey \& Cole 1993):\\
 The linear overdensity computed at a given
point of an initial snapshot of the Universe fluctuates  when the
smoothing scale decreases. This  fluctuation is a Markovian
process when the smoothing is performed using a top-hat window in Fourier
space. For any value of the smoothing scale $R$ the overdensity
field is assumed to be Gaussian. The mass $M$ contained in a given
scale $R$ depends on the window function used. For a top-hat
window the relation is:  $M=\frac{4}{3}\rho_{b,i}\
R^3=\frac{\Omega_{m,i} H^2_i}{2G}R^3$, where $\rho_{b,i}$ and
$\Omega_{m,i}$ are the values of the mean density and the density
parameter of the Universe, $G$  is the gravitational constant  and
$H_i$ is the Hubble's constant. The index $i$ indicates that all
the above values are calculated at the initial snapshot. Mass
dispersion $\sigma^2$ at scale $R$ is a function of mass $M$ and
is usually denoted by $S$, that is $S(M)\equiv \sigma^2[R(M)]$.
Let the random walk of the overdensity cross for first time a
given barrier $B(S,z)$ at some value $S_0$ of $S$. The mass
element associated with the random walk is considered to belong to
a halo of mass $M_0=S^{-1}(S_0)$ at the epoch with redshift $z$.
The distribution of haloes, at some epoch $z$, is related to the
first
crossing distributions, by the random walks, of the barrier that corresponds to epoch $z$.\\
The simplest form of the barrier comes from the spherical model.
It is well known that in an Einstein-de Sitter Universe, a
spherical overdensity collapses at $z$ if the linear extrapolation
of its value up to the present exceeds $\delta_{sc}\approx 1.686$,
and this value provides a first reasonable choice for the barrier.
The involved quantities (density overdensities, dispersion) are
usually extrapolated to the present epoch and thus the barrier in
the spherical collapse model is written in the form
$B(S,z)=1.686/D(z)$, where $D(z)$ is the growth factor derived by
the linear theory with $D(z=0)=1$. The form of the spherical
barrier permits the
analytical evaluation of the first crossing distribution $f(S)$. \\
Despite the simplicity of the spherical model, its results agree
relatively well with the results of N-body simulations (e.g.
White, Efstathiou \& Frenk 1993; Lacey \& Cole 1994; Gelb \&
Bertschinger 1994; Bond \& Myers 1996). Deviations appear in the
resulting mass functions at both high and low masses. Sheth \&
Tormen \cite{sheth} consider a barrier of the form
$B(S,z)=p(z)+q(z)S^\gamma$ with $p(z)=0.840\delta_c(z),
q(z)=0.505{\delta_c(z)}^{-0.23}$, where $\delta_c(z)=1.686/D(z)$
and $\gamma=0.615$, in order to describe the ellipsoidal form of
collapse.  The first crossing distribution $f(S,z)$ that results
from the ellipsoidal model works better than the spherical one.
For example, Yahagi et al. \cite{yah} showed that the multiplicity
function resulting from N-body simulations is far from the
predictions of spherical model while it shows an excellent
agreement with the results of the ellipsoidal model. On the other
hand, Lin et al. \cite{lin} compared the distribution of formation
times of haloes formed in N-body simulations with the formation
times of haloes formed in terms of the spherical collapse model.
They found that N-body simulations give smaller formation times.
Hiotelis \& del Popolo \cite{hiot-del} used merger trees to show
that the ellipsoidal collapse model leads to formation times that
are shifted to smaller values relative to the spherical collapse
model. Thus, a better agreement with the predictions of N-body
simulations is achieved. The distributions of formation times are
studied in more detail by other authors (Giocoli et al. 2007). \
We note that in the cases of a barrier with $\gamma=1$, or
$\gamma=0.5$, the corresponding first crossing distribution can be
found analytically (e.g. Mahmood \& Rajesh 2005). For the above
non-linear barrier, Sheth \& Tormen \cite{st02} proposed an
analytical expression. It is shown  that this analytical
expression approximates well the exact expression found as a
numerical
solution of an integral equation (Zhang \& Hui 2008 ).\\
The constrained first crossing distribution is given by the
relation:
\begin{eqnarray}
f(S,z/S,z_0)\mathrm{d}S=\frac{1}{\sqrt{2\pi}}\frac{|T(S,z/S_0,z_0)|}{(\Delta
S)^{3/2}}\nonumber\\
\exp\left[-\frac{(\Delta B)^2}{2\Delta S}\right]\mathrm{d}S
\end{eqnarray}
where $\Delta S=S-S_0$, $\Delta B=B(S,z)-B(S_0,z_0)$ and the
function $T$ is given by:
\begin{eqnarray}
T(S,z/S_0,z_0)=B(S,z)-B(S_0,z_0)+\nonumber\\
\sum_{n=1}^5\frac{[S_0-S]^n}{n!}\frac{\partial ^n}{\partial
S^n}B(S,z)
\end{eqnarray}
Given that a mass element is a part of a halo of mass $M_0$ at
redshift $z_0$ the probability that at higher redshift $z$ this
mass element was a part of a smaller halo $M$ is given by Eq.(1).
The unconstrained expression, $f(S,z)$, results by setting
$S_0=B(S_0,z_0)=0$. The analytical expression is very useful since
it allows  the construction of merger trees.\\
The purpose of this paper is:
\begin{enumerate}
\item To construct merger trees able to give merger rates of dark
matter haloes.
\item To compare these merger rates with those predicted by the results
of N-body simulations.
\item To extend the calculations to scales that are not accessible
by numerical simulations and finally,
\item to study the role of some of the main parameters involved.
\end{enumerate}
In Sect. 2 we give a brief description of the tree code used.
Then, the definition of merger rates is presented and the
analytical formulae, predicted by other authors from the results of
N-body simulations, are given.\\
In Sect. 3 we present our results.
\section{Tree construction, definitions of halo merger rates fitting formulae}
\subsection{Tree  construction}
Merger-trees used in this paper are constructed using Eq.(1). Let
us assume a descendant  halo of mass $M_d$ at redshift $z_d$. We
study its past by the following procedure: A new larger redshift
$z_p$ is chosen. This is done by solving for $z_p$ the equation
$\delta_c(z_p)-\delta_c(z_d)=D$, where $D$ is a constant (one of
the parameters of the algorithm). Then, a value $S_p$ is chosen
from the distribution (1). The mass of the progenitor is
$M_p=S^{-1}(S_p)$. This progenitor is accepted if its mass is
larger than a lower limit $M_{min}$ and smaller than the mass left
to be resolved. If $M_p$ is less than $M_{min}$ then $M_p$ is
added to a sum that is named $M_{accr}$. The mass left to be
resolved is, at the choice of the k-th progenitor,
$M_{left}=M_d-\sum_{l=1}^{l=k-1}M_{p,l}-M_{accr}$. If $M_{left}$
is larger than $M_{min}$ we proceed to the selection of a next
progenitor, while if $M_{left}$ is smaller than $M_{min}$ we
proceed to the next redshift. It is obvious that in such
construction, the number of progenitors can be larger than two,
despite the original assumption of Lacey \& Cole \cite{lc93}, and
this leads to a better representation of the distribution of
progenitors. Our algorithm  is based on the  ``N-branch'' idea of
Somerville \& Kollat  \cite{som}, but  extended in order to
incorporate aspects of the ellipsoidal collapse results. A
complete description of the tree construction is given in Hiotelis
\& del Popolo \cite{hiot-del}. The comparisons between the
predictions of the tree and analytical predictions of the
distribution of the number of progenitors show that the tree
method is reliable in following the evolution of structures.
 Various  tree-construction algorithms have been presented in the
 literature (Cole 1991; Kauffmann \& White 1993; Sheth \& Lemson 1999; Cole et al. 2000;
  Neinstein \& Dekel  2008). The accuracy of these algorithms is usually against simplicity. For example,
  the algorithm of Neinstein \& Dekel  \cite{nei} requires the solution of several differential equations
   with nontrivial boundary conditions.\\
\subsection{Definition of merger-rates}
We examine one descendant halo from a sample of $N_d$ haloes with
masses in the range $M_d, M_d+\mathrm{d}M_d$  present at redshift
$z_d$. For a single halo the procedure is as follows: Let
$M_{p,1},M_{p,2}...M_{p,k}$ be the masses of its $k$ progenitors
at redshift $z_p > z_d$. For matter of simplicity we assume that
the most massive progenitor is $M_{p,1}$. We define
$\xi_i=M_{p,i}/M_{p,1}$ for $i\geq 2$ and we assume that the
descendant halo is formed by the following procedure: During the
interval $dz=z_p-z_d$ every one of the progenitors with $i\geq 2$
merge with the most massive progenitor $i=1$ and form the
descendant halo we examine.\
 We repeat the above procedure for all
haloes in the range $M_d, M_d+\mathrm{d}M_d$ found in a volume $V$
of the Universe. Then, we find the number denoted by $N$ of all
progenitors with $\xi_i,  i\geq 2$ in the range $(\xi,
\xi+\mathrm{d}\xi)$ and we calculate the ratio
$N/(V\mathrm{d}z\mathrm{d}M_d\mathrm{d}\xi)$. We define the merger
rate $B_m$ as follows:
\begin{equation}
B_m(M_d,\xi,z_p:z_d)=\frac{N}{V\mathrm{d}z\mathrm{d}M_d\mathrm{d}\xi}
\end{equation}
Let the number density of haloes  with masses in the range $M_d,
M_d+\mathrm{d}M_d$ at $z_d$ be
$n(M_d,z_d)=\frac{N_d(M_d,z_d)}{V\mathrm{d}M_d}$.  The ratio
$B_m/n=N/(N_d\mathrm{d}z\mathrm{d}\xi)$ measures the mean number
of mergers per halo, per unit redshift, for descendant haloes in
the range
$M_d, M_d+\mathrm{d}M_d$ with progenitor mass ratio $\xi$.\\
We note that the definition of mean merger rate is exactly the
same as in Fakhouri \& Ma \cite{FM08} (FM08 hereafter). We also
use the assumption that all progenitors merge with the most
massive one (see FM08 for
a discussion of this assumption.)\\
 Lacey \& Cole \cite{lc93} showed that in the spherical model the transition
rate is given by:
\begin{eqnarray}
  r(M\longrightarrow M_d/z_d)\mathrm{d}M_d=
  \left(\frac{2}{\pi}\right)^{1/2}\left[\frac{\mathrm{d}\delta_c(z)}{\mathrm{d}z}\right]_{_{z=z_d}}\nonumber\\
  \times \frac{1}{\sigma^2(M_d)}
  \left[1-\frac{\sigma^2(M_d)}{\sigma^2(M)}\right]^{-3/2}\left[\frac{\mathrm{d}\sigma(M)}{\mathrm{d}M}\right]_{_{M=M_d}}\nonumber\\
  \times\exp\left[-\frac{\delta^2_c(t)}{2}
  \left(\frac{1}{\sigma^2(M_d)}-\frac{1}{\sigma^2(M)}\right)\right]\mathrm{d}M_d
  \end{eqnarray}
  This provides the fraction of the mass belonging  to haloes of mass $M$ that merge
  instantaneously to form haloes of mass in the range $M_d, M_d+\mathrm{d}M_d$ at
  $z_d$. The product $r\cdot f_{sc}(M,z_d)\mathrm{d}M$, where $f_{sc}(M,z)$ is the unconditional first crossing
  distribution for the spherical model, gives the above fraction of mass as a fraction of the
  total mass of the  Universe and successively multiplying by $(\rho_b/M)\cdot V$  the number
  of those haloes is found. Then, by dividing by
  $(\rho_b/M_d)\cdot V\cdot f_{sc}(M_d,z_d)\mathrm{d}M_d$ (that equals
  to the number of the descendant haloes) we find:
  \begin{eqnarray}
  \frac{N}{N_d\mathrm{d}z}=\sqrt{\frac{2}{\pi}}
  \frac{M_d}{M}\frac{1}{\sigma^2(M)}\frac{\mathrm{d}\sigma(M)}
  {\mathrm{d}M}\left[\frac{\mathrm{d}\delta_c(z)}{\mathrm{d}z}\right]_{_{z=z_d}}\nonumber\\
  \times \left[1-\frac{\sigma^2(M_d)}{\sigma^2(M)}\right]^{-3/2}\mathrm{d}{M}
  \end{eqnarray}
  Assuming a binary merge, where $\xi$ is the ratio of the small
  progenitor to the large one ($\xi=(M_d-M)/M$), using $\mathrm{d}M=\frac{M^2}{M_d}
  \mathrm{d}\xi$ and substituting in (5) we have the final
  expression for the binary spherical case, that is:
  \begin{eqnarray}
  \frac{B_m}{n}= \frac{N}{N_d\mathrm{d}z\mathrm{d}\xi}=\sqrt{\frac{2}{\pi}}
  \frac{M}{\sigma^2(M)}\frac{\mathrm{d}\sigma(M)}{\mathrm{d}M}
  \left[\frac{\mathrm{d}\delta_c(z)}{\mathrm{d}z}\right]_{_{z=z_d}}\nonumber\\
  \times\left[1-\frac{\sigma^2(M_d)}{\sigma^2(M)}\right]^{-3/2}
  \end{eqnarray}
\subsection{Fitting formulae}
FM08 analyzed the results of the Millennium simulation of Springel
et al. \cite{spri}. Stewart et al. \cite{stew}, (SBBW08
hereafter), used their high-resolution N-body simulations to study
the merger rates of dark matter haloes.  Fitting formulae proposed
by the above authors are separable in the three variables, mass
$M_d$, progenitor ratio $\xi$ and redshift
$z$. These formulae are of the form:\\
\begin{equation}
\frac{B_(M_d,\xi,z_p:z_d)}{n(M_d,z)}=A\cdot F(M_d)G(\xi)H(z)
\end{equation}
FM08 proposed $A=0.0289,
F(M_d)=\left(\frac{M_d}{\tilde{M}}\right)^{a_1},G(\xi)={\xi}^{a_2}\exp\left[\left(\frac{\xi}{\tilde{\xi}}\right)^{a_3}\right],
H(z)=\left(\frac{\mathrm{d}\delta_c}{\mathrm{d}z}\right)^{a_4}_{_{z=z_d}}$.
where the values of the parameters are
$\tilde{M}=1.2\times10^{12}M_{\sun}, A=0.0289, \tilde{\xi}=0.098,
a_1=0.083, a_2=-2.01, a_3=0.409, a_4=0.371$.\\
On the other hand, SBBW08 proposed: $A=0.27,
F(M_d)=\left(\frac{M_d}{\tilde{M}}\right)^{b_1},G(\xi)=(1-\xi)^{b_3-1}[(b_3-b_2)\xi+b_2]/{\xi}^{b_2+1},
H(z)=\left(\frac{\mathrm{d}\delta_c}{\mathrm{d}z}\right)^{b_4}_{_{z=z_d}}$
where $\tilde{M}=10^{12}h^{-1}M_{\sun}, b_1=0.15, b_2=0.5,b_3=1.3,
b_4=2.$\\
Note that the formulae proposed by the above authors show some
significant differences. First, the dependence on redshift $z$
differs in the above two formulae. Although the quantity
$\mathrm{d}\delta_c/\mathrm{d}z$ does not vary significantly with
redshift, the exponents $a_4$ and $b_4$ can cause significant
differences in the merger rates. Second, exponents $a_1$ and $b_1$
that define the dependence on the mass are quite different. In the
 approximation of FM08 mean merger rates are practically
 independent on mass.\\
 Additionally, the above formulae show significant differences in their slope at small and large
  values of $\xi$. The logarithmic slope of $G$,
  $\mathrm{d}\ln G(\xi)/\mathrm{d}\ln \xi$, at $\xi =0$ is $a_2=-2.01$ for the FM08 model, and
  $-b_2-1=-1.5$ for the SBBW08 model. For $\xi \rightarrow 1$ the
  above logarithmic slope is $a_2+a_3(1
  /\tilde{\xi})^{a_3}=-0.952$ for the FM08 model and it tends to
  $-\infty $ in the formula of SBBW08.\\
   Therefore, it is interesting to study merger rates predicted by merger-trees and to compare their characteristics with
   those of the above fitting formulae.
\section{Results}
\begin{figure}
\includegraphics[width=9cm]{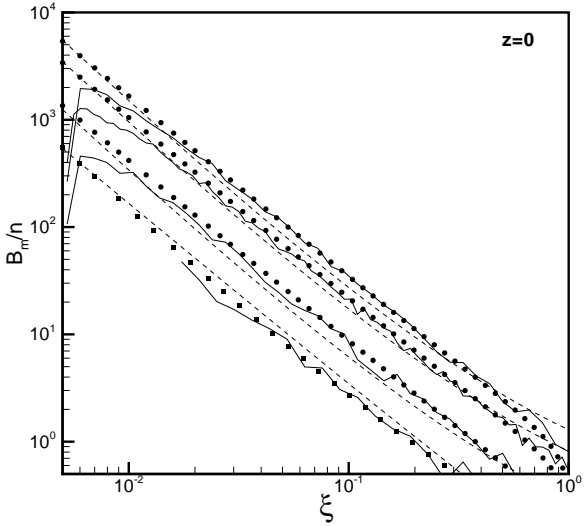}
 \caption{From top to bottom, dots correspond to the mean present merger rates ($z=0$) for haloes with
 present day masses $10^{15}h^{-1}M_{\sun}$, $10^{14}h^{-1}M_{\sun}$, $10^{12}h^{-1}M_{\sun}$ and $10^{10}h^{-1}M_{\sun}$
 respectively, predicted
 by the formula proposed by FM08 for $a_1=0.2$ and
 $a_4=1.1$. Dashed lines are predicted by the formula proposed by SBBW08 for $b_1=0.2$ and $b_2=0.7$.
 Solid lines are the predictions of merger trees described in the text by Eq.(3)  for $z_d=0$ and $z_p=0.0556$.
 We used a sample of $10000$ present-day haloes for each of the three cases and we evolved the system by a single
 time-step for $D=0.05$. The minimum mass $M_{min}$ in every case equals to 0.005 times $M_{d}.$}
 \label{mnfig1}
\end{figure}
We used a  flat model for the Universe with $\Omega_{m,0}=0.3$ and
$\Omega_{\Lambda ,0}=0.7$ and a  power spectrum proposed by Smith
et al. \cite{smith}
 given by:
\begin{equation}
P(k) =\frac{Ak^n}{[1+a_1k^{1/2}+a_2k+a_3k^{3/2}+a_4k^2]^b}
\end{equation}
The values for the parameters are: $n=1,~ a_1=-1.5598,~
a_2=47.986, ~a_3=117.77,~ a_4=321.92$ and $b=1.8606$. Smoothed
fields are calculated using the top-hat window function. The
constant $A$ of proportionality, is found using the procedure of
normalization. We used two different values for the normalization
namely $\sigma_8\equiv \sigma(R=8h^{-1}\rmn{Mpc})=0.9$ and $1$
respectively. We also use a system of units where
$M_\rmn{unit}=10^{12}h^{-1}\rmn{M_{\sun}},~
 R_\rmn{unit}=h^{-1}\rmn{Mpc} $ and $t_\rmn{unit}=1.515\times10^{7}h^{-1}\rmn{years}$.
  In this system of units,
$H_0/H_{\rmn{unit}}=1.5276$.
\begin{figure}
\includegraphics[width=9cm]{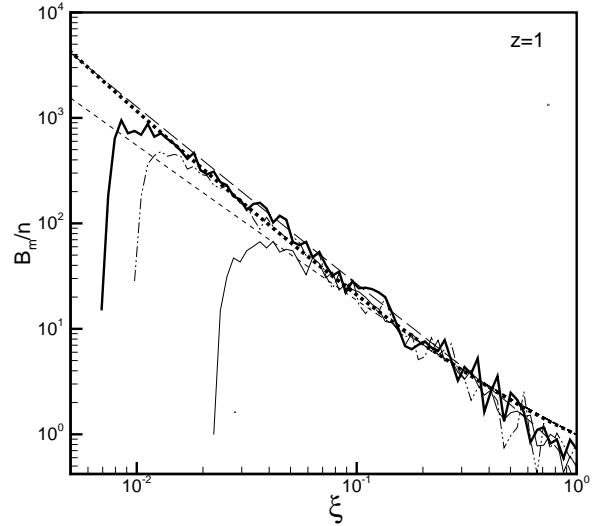}
\caption{Thin solid line, dashed-dot-dot line and thick solid line show the mean merger rate at $z\approx 1$ for haloes
 with masses in the range  $M_1=1.5-2.5\times
10^{13}h^{-1}M_{\sun}, M_2=4.5-5.5\times
10^{13}h^{-1}M_{\sun}$ and $M_3=6.5-7.5\times 10^{13}h^{-1}M_{\sun}$,
respectively. The merger tree used started at $z=0$ with $D=0.05$
for a sample of $36500$ haloes each of mass
$10^{14}h^{-1}M_{\sun}$. After 22 time steps the redshifts are
$z_d=0.979, z_p=1.021$. The resolution mass is $M_{min}=5\times
10^{11}h^{-1}M_{\sun}$. Long dashes, big dots and small dashes are
the results from the formulae of SBBW08, FM08 and the binary
spherical model for $M_d=5\times 10^{13}h^{-1}M_{\sun}$ at $z=1$.}
\end{figure}
We performed a large number of tree realizations in order to study
merger rates. First, we found that good fits are achieved by both
of the above formulae but for different values of the parameters
than those proposed by the above authors. We found that merger
rates depend on the mass of the descendant halo as ${M_d}^{0.2}$
and on the redshift as $[\mathrm{d}\delta_c/\mathrm{d}z]^{1.1}$.
So, in the comparisons that follow, the formulae of Eq. 7 are used
with $a_1=b_1=0.2, a_4=b_4=1.1$. Additionally, in the formula of
SBBW08, we use the value of $b_2=0.7$ instead of $b_2=0.5$.\\
In Fig.1, comparisons between the predictions of merger trees and those of
formulae given by Eq. 7 are shown. The power spectrum used is that
given by Eq. 8 for $\sigma_8=1$. Details are given in
the caption of the figure. We note a very good agreement but also a rapid fall of
the predictions of merger trees for small values of $\xi$. We will
show below that this is a matter of resolution that depends on the value of $M_{min}$.\\
The predictions of this figure are just after  one time step of the
tree algorithm. For a higher redshift  $z_1$ one has to use one of
the two alternatives: To start with a sample of haloes at $z_1$
and after a single time step to move to a new redshift $z_{p1}$, to
calculate the merger rates, or to start with a sample of haloes at
the present epoch $z=0$ and to calculate merger rates after a
number of time steps when the redshift has a desirable value.
\begin{figure}
\includegraphics[width=9cm]{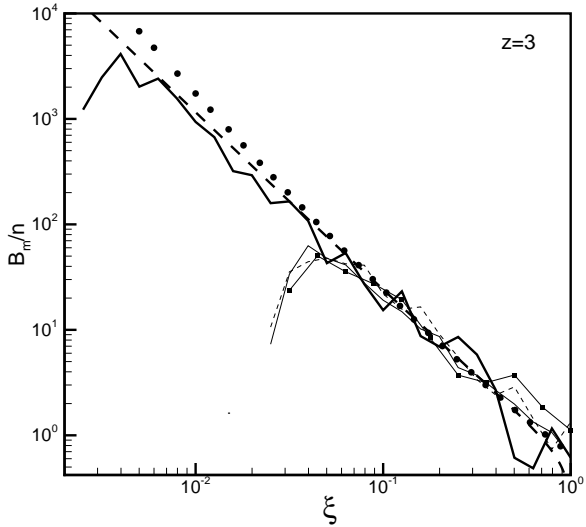}
 \caption{ Merger rates for descendant haloes
    with masses in the range $1.5-2.5\times  10^{13}h^{-1}M_{\sun}$, for an initial sample of 10000 haloes each with mass
  $10^{14}h^{-1}M_{\sun}$ after 72 steps (thin solid line), after 144 steps (thin solid line with large dots) and after
  36 steps (line with thin and small dashes). These three lines correspond to about the same redshift (z=3), and to the
  same resolution mass $M_{min}=5 \times 10^{11}h^{-1}M_{\sun}$.Thick solid line
  is the prediction for smaller  minimum  mass, $M_{min}=5 \times 10^{10}h^{-1}M_{\sun}$. Big dots are the predictions
  of FM08, while large thick dashes, the predictions of SBBW08. }
  \label{rfig3}
\end{figure}
Obviously, the second approach tests the ability of the merger
tree to follow the time evolution of structures and it is close
to the nature of  N-body simulations. This approach is followed
in our calculations.\\
Fig.2 shows merger rates predicted by a merger tree, started at $z=0$, after
22 time steps, for $D=0.05$ and  $\sigma_8=1$. Redshifts are
$z_d=0.979, z_p=1.021$. The sample of haloes at $z=0$ consists of
$36500$ haloes each with mass $M_0=10^{14}h^{-1}M_{\sun}$. The
number of haloes  in the range $1.5-2.5\times
10^{13}h^{-1}M_{\sun}$ at $z_d$ is $N_p=19241 $ and $N_d=15465$.
In the range $4.5-5.5\times 10^{13}h^{-1}M_{\sun}$ there are
$N_p=7587$ and $N_d=4961$ haloes. Finally, in the range
$6.5-7.5\times 10^{13}h^{-1}M_{\sun}$ there are $N_p=5803$ and
$N_d=3330$ haloes. The resolution mass, $M_{min}$ used is $5\times
10^{11}h^{-1}M_{\sun}$. For $z=1$ long dashes show the formula of
SBBW08,
 dots the formula of FM08 and small dashes the prediction of the
 binary spherical model given by Eq. 6.
In the following we  examine the role of some -of the large
number- of the parameters involved in the construction of  merger
trees. These are:
 \begin{enumerate}
 \item
  the resolution mass $M_{min}$
 \item the step in redshift, defined by the parameter $D$
 \item the value of $\sigma_8$ and
 \item the number of realizations, that is
 the number of haloes at $z=0$.
 \end{enumerate}
   Resolution mass $M_{min}$ is a crucial parameter. A sample of
 present day haloes of mass $M_0$ is analyzed to smaller and smaller haloes as the redshift becomes higher.
 Let $M_d$ be the mass of a descendant halo at some high redshift, $M_1$ its largest progenitor and  $M_2$
 another progenitor. Obviously, $M_1 \leq M_d-M_{min}$ and $M_2 \geq M_{min}$.
  Thus $\xi=\frac{M_2}{M_1}\geq \frac{M_{min}}
 {M_d-M{min}}\equiv\xi_{min}$ and consequently  ${\xi}_{min}$ and $M_{min}$ are related by $M_{min}=M_d(1+{\xi}_{min}^{-1})^{-1}$.
  Since at high redshifts $M_d$ is significantly smaller than $M_0$,  the condition for
  the merger rate curves to extend to the left up to values as small as   ${\xi}_{min}$ is
  $M_{min}\ll M_0(1+{\xi}_{min}^{-1})^{-1}$. We found that for decreasing $M_{min}$ the value $\xi$, at which
  the merger rates curves show their rapid fall, moves to the left.
  Thus, this rapid fall seen in Figs 1 and 2 is clearly a matter
  of resolution.\\
    On the other hand,  the results do not seem to be sensitive to the values of the step in redshift $z$. This step
  depends on the  parameter D described in section 2.1. We examined cases, for D=0.025, D=0.05 and D=0.1.
   We found that the results are the same.\\
 Fig.3  shows the role of the above two parameters. It presents merger rates, for descendant haloes
    with masses in the range $1.5-2.5\times  10^{13}h^{-1}M_{\sun}$, for an initial sample of 10000 haloes, each with mass
  $10^{14}h^{-1}M_{\sun}$, after 72 steps (thin solid line), after 144 steps (thin solid line with large dots) and after
  36 steps (line with thin and small dashes). All lines correspond to about the same redshift (z=3), since we used different values for the time step parameter,
  $D=0.05$, $D=0.025$ and $D=0.1$, respectively. The corresponding redshifts are $(z_d=2.971,
z_p=3.011),~(z_d=3.005,z_p=3.025),~(z_d=2.921,z_p=3.000)$, respectively.
       We see that the resulting merger rates are similar, although the values of $dz$ used in Eq. 3
  differ, $dz\approx 0.02, 0.04, 0.08$.\\
\begin{figure}
\includegraphics[width=9cm]{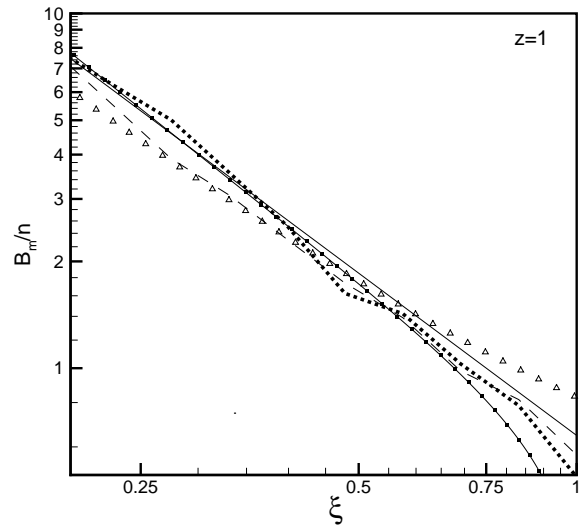}
 \caption{Merger rates for descendant haloes
    with masses in the range $1.5-2.5\times  10^{13}h^{-1}M_{\sun}$ at $z=1$. Dots are the predictions
     for the power spectrum normalized to $\sigma_8=1$, while dashes
     correspond to $\sigma_8=0.9$. Solid line  shows a law $\propto \xi^{-1.52}$ for comparison.
     Deltas correspond to the model of FM08
     while solid line with black squares to the model of SBBW08.}
  \label{rfig4}
\end{figure}
In the same figure the thick solid line is the prediction for a resolution mass an order of magnitude smaller,
  namely $M_{min}=5 \times 10^{10}h^{-1}M_{\sun}$ and shows the
  above described role of $M_{min}$. Finally, big dots and thick dashes are the predictions
  of FM08 and SBBW08, respectively, for values of the parameters given above.
  All calculations in Fig.3 are for $10000$ present day haloes and for $\sigma_8=1$.
  The above described role of the two parameters is the same for the range of present day masses, $10^{10}M_{\sun}\leq M_0
\leq10^{15}M_{\sun}$, we examined. We note that small values of
$M_{min}$ lead to large numbers of haloes at the past. For
example, $10000$ present day haloes with mass $10^{15}M_{\sun}$
have for $M_{min}=5\times 10^{12}M_{\sun}$, about $1.2\times 10^5$
progenitors at $z \approx 3$. For the minimum value of $M_{min}$
that we have used, that is $M_{min}=5\times 10^{10}M_{\sun}$ and
corresponds to a resolution $1:20000$, at the same redshift, the
number of progenitors is larger than $1.5\times 10^6$. Thus, tree
construction becomes a computing time consuming
procedure.\\
Major mergers seem to play an important role in the formation of
dark matter haloes. N-body simulations show that haloes
which experienced a recent major merger event, appear to have lower
concentrations and steeper inner density profiles (e.g. Ascasibar
et al. 2003, Tasitsiomi et al. 2004) as well as larger values of
spin parameters (Gardner 2001, Peirani et al. 2004). This last
result is also supported by semi-numerical results (e.g. Vitvitska
et al. 2002, Hiotelis 2008). Thus, it is interesting to study in
more detail the behavior of merger rates curves at large values
of $\xi$. This demands smooth curves and consequently large number
of haloes. So, in Fig.4 we present an example of the role of the
last two of the parameters we examined. These parameters are the
number of present day haloes and the value of $\sigma_8$. All
curves of this figure correspond to $z=1$. Dots are the predictions
for a set of $36500$ present day haloes each with mass $M_0=
10^{14}h^{-1}M_{\sun}$ for a power spectrum with $\sigma_8=1$ and
for descendant haloes in the range $1.5\times
10^{13}h^{-1}M_{\sun}- 2.5\times 10^{13}h^{-1}M_{\sun}$. At $z=1$,
there are $N_p=19241$ and $N_d=15456$ haloes. Dashes are the
predictions for a set of $1.5\times 10^5$ present day haloes of
the same mass for $\sigma_8=0.9$ and for the above range of mass
of the descendant haloes. At $z=1$, there are $N_p=104873$ and
$N_d=70276$ haloes. Solid line shows a law $\propto \xi^{-1.52}$
for comparison. Deltas are the predictions of the fitting formula
of FM08 and finally, the solid line with big dots shows the predictions
of SBBW08. Fig.5 is similar to Fig.4 but for $z=3$ and for
descendant haloes in the range $5\times 10^{12}h^{-1}M_{\sun}-
10^{13}h^{-1}M_{\sun}$. There are $N_p=26610$ and $N_d=24540$ for
the tree corresponding to the dashed line while there are
$N_p=129074$ and $N_d=109678$ for the tree corresponding to
dots. It is clear from the above two figures, that - at the level
of accuracy of the calculations of this paper - no differences can
be detected due to the difference between $0.9$ and $1.0$ of the
values of $\sigma_8$. Larger number of present day haloes leads,
obviously, to smoother curves but they cause no difference in the overall shape. \\
\section{Discussion}
The above study of merger rates that are predicted by merger trees
based on the extended PS theory using the ellipsoidal collapse
model, leads to the following conclusions that hold for haloes in the range of mass
$10^{12}h^{-1}M_{\sun} - 10^{15}h^{-1}M_{\sun}$
 and for redshift $0\leq z\leq 3$.
\begin{figure}
\includegraphics[width=9cm]{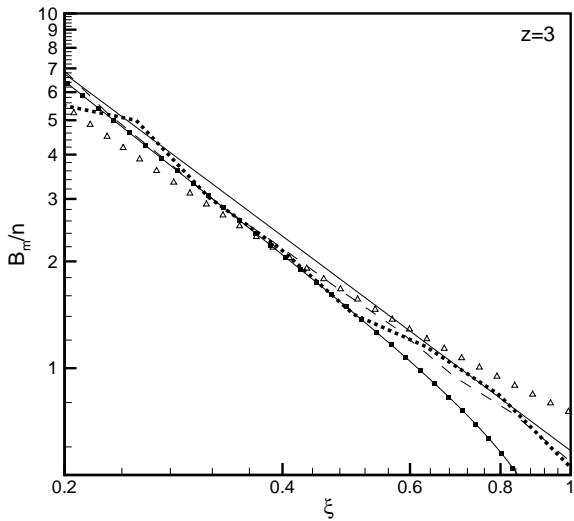}
 \caption{As in Fig.4 but for z=3.}
  \label{rfig5}
\end{figure}
\begin{enumerate}
  \item Merger rates depend on the mass of the descendant halo as $\sim M^n$ with n$=0.2$. This
  dependence is in practice indistinguishable from the value $n=0.15$ proposed from SBBW08 but is
   far from the value of $n=0.083$ proposed by  FM08.
  \item We found that merger rates depend on the redshift through the quantity
  $[\mathrm{d}\delta_c(z)/\mathrm{d}z]^l$, where  $l=1.1$. This
  value of the exponent $l$ is close to the value predicted by the binary spherical case (see Eq.6) and between the values
  0.371 and 2 proposed by FM08 and SBBW08, respectively.
  \item The results of merger trees are in better agreement (for the
  above values of $n$ and $l$) with the predictions of formulae of FM08 and SBBW08
  than with the predictions of binary spherical model given by
  Eq.6. The binary spherical model underestimates merger rates for small values of $\xi$ while merger trees, for proper resolution,
  give results much closer to those of N-body simulations.
  \item  For large values of $\xi$, merger rate curves
  scale approximately as $ \approx {\xi}^{-1.5}$.
   \item We examined steps in redshift from $\approx 0.02$ to $\approx 0.08$ and we found no differences in the results.
   \item Smaller values of the resolution mass $M_{min}$ give a better agreement with
    the above fitting formulae for small values of $\xi$.
    \end{enumerate}
The construction of reliable merger-trees is a subject under
current  investigation. These algorithms usually have fundamental
problems  regarding mass conservation, accurate representation of
the distribution of progenitors etc.
    As regards N-body simulations, in addition to their resolution problems, it is characteristic that some of
    their results   depend on the techniques used for their derivation. For example, Bett et al. (2007) showed that the
    values of spin
    parameters of haloes and their behavior as a function of mass
    depends crucially on the halo-finding algorithm. However, the study of physical  parameters -as for example a
    more accurate power spectrum- seems not
    to be proper at this stage. Probably, the results are more sensitive to the method used for their analysis than
    to physical parameters and thus, any detailed description between the results of N-body and those of
    merger-trees may not be very useful. However, there are characteristic trends that show interesting
    agreement.\\
    Further improvements of tree construction algorithms as well of
the quality of N-body simulations
     could help to understand better the physical picture during
  the process of the formation of dark matter haloes.

\section{Acknowledgements} I would like to thank the
\emph{Empirikion Foundation} for its financial support.
%\end{acknowledgements}

\end{document}